# SIDE, a fiber fed spectrograph for the 10.4 m Gran Telescopio Canarias (GTC)

F. Prada[*a], M. Azzaro[a], O. Rabaza[a], J. Sánchez[a], M. Ubierna[a]

[a] Instituto de Astrofísica de Andalucía (IAA-CSIC), C/ Camino Bajo de Huétor 50, 18008 Granada, Spain

**ABSTRACT**

SIDE (Super Ifu Deployable Experiment) will be a second-generation, common-user instrument for the Grantecan (GTC) on La Palma (Canary Islands, Spain). It is being proposed as a spectrograph of low and intermediate resolution, highly efficient in multi-object spectroscopy and 3D spectroscopy. SIDE will feature the unique possibility of performing simultaneous visible and IR observations for selected ranges. The SIDE project is leaded by the Instituto de Astrofsica de Andaluca in Granada (Spain) and the SIDE Consortium is formed by a total of 10 institutions from Spain, Mexico and USA. The feasibility study has been completed and currently the project is under revision by the GTC project office.

## 1. INTRODUCTION

The SIDE spectrograph will provide low resolution (R = 1500, 4000, 5000) and high resolution (R = 8000, 15000, 29000). The low resolution spectrograph is called "Dual VIS-NIR" because it will be capable of simultaneous observations in the Visible and near Infrared bands (0.4 – 1.7 μ) by using a dichroic. The high resolution device is called Hi-Res VIS spectrograph as it works in the Visible band only (0.35 – 0.87 μ). SIDE will work in three observing modes: MOS, SIFU and mIFU. The Dual VIS-NIRspectrograph will work in all three modes (MOS, SIFU and mIFU), while the Hi-Res VIS will be fed only by MOS and mIFU units. Due to the large number of fibers, the Dual VIS-NIRspectrograph will be a set of ~10 identical spectrographs, located, together with the Hi-Res spectrograph in a dedicated and conditioned room mounted on the telescope structure. The basic parameters of SIDE are shown in Table 1.

Table 1. A resume of the basic parameters of SIDE.

| Location | GTC 10.4m Nasmyth focus |
|---|---|
| FOV | MOS: 20' diameter<br>mIFU: 3"x 3" each over an 8' focal plane SIFU 30"x 30" |
| Instrument observing modes | MOS, SIFU, mIFU |
| N. of science units | ~1000 MOS<br>~2500 fibers in SIFU ~27 mIFUs of 36 fibers each |
| Resolution, spectral coverage and mode | 1500 < R < 30000<br>spectral window: 350-1700 nm<br>Nod&Read observing mode |

---

[*] fprada@iaa.es

The preferred location for the MOS units, configured by a Fiber Positioner Robot, is the Nasmyth focus of GTC, with 20' of FOV and some 950 science units. The preferred location for the SIFU and mIFU units is the Folded Cassegrain focus, with about 2500 fibers for the SIFU and about 27 mIFUs.

The goal of SIDE is to provide the GTC community with new and unique observing capabilities, i.e.

- Intermediate spectral resolution
- Survey spectroscopy
- 3D spectroscopy

**1.1 SIDE schematics**

Fig. 1 shows the basic flow of light through the instrument: light enters the instrument at the two foci behind the A&G boxes (Nasmyth and Folded Cassegrain) for MOS and mIFUs and SIFU respectively.

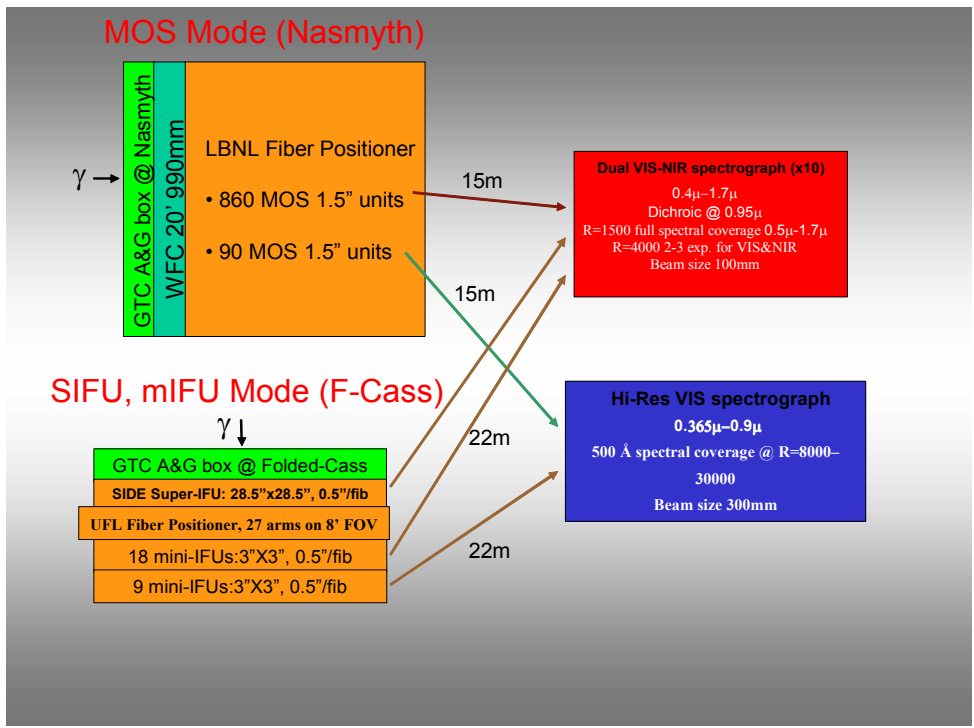

Fig. 1. Basic flow of light through the instrument.

**1.2 SIDE Consortium**

The project is leaded by the Instituto de Astrofísica de Andalucía (IAA-CSIC), based in Granada, Spain. After the project received support from the GTC SAC, a Science Team and an Instrument Definition Team were formed and the basis of the instrument were defined. The Feasibility Study phase, which started the 1st of December of 2006 and finished on 1st December 2007, is funded by the Spanish MEC and the GTC, together with the Consortium members support.

After the first kick-off meeting in Sevilla (Sept. 2006), the SIDE Consortium was formed and several institutions from Spain, Mexico and USA committed themselves for the project. At present, 10 Institutions with worldwide recognized prestige form the SIDE Consortium:

1. Instituto de Astrofísica de Andalucía (IAA-CSIC)
2. Instituto de Astrofísica de Canarias (IAC)
3. Institut d'Estudis Espacials de Catalunya (IEEC)
4. Institut de Física d'Altes Energies (IFAE)
5. Instituto Nacional de Astrofísica de Óptica y Electrónica (INAOE)
6. Universitat de Barcelona (UB)
7. Universidad Nacional Autónoma de México (UNAM)
8. Universidad Complutense de Madrid (UCM)
9. University of Florida (UFL)
10. Lawrence Berkeley National Laboratory (LBNL)

## 2. SCIENCE CASE

### 2.1 Science with SIDE

A great variety of science cases could be addressed with SIDE, covering from Cosmology, Stellar Physics, Solar System and Extragalactic Astronomy of which some are reported in the list given here.

Science in MOS mode:
- Galaxy redshift surveys: large scale structure, dark energy, galaxy formation
- Galaxy clusters, AGNs, QSOs and IGM at high z
- Galactic structure and evolution: radial velocities and metallicities (bulge, disc, globular clusters and tidal streams)
- Target follow-up of SDSS, ALHAMBRA, COMBO-17/GEMS, VISTA, ….
- Stellar populations and star formation rates of massive galaxy samples
- dSph, M31, M33 and Local Group studies

Science in IFU mode:
- Ionized gas in galaxies
- Jet kinematics and physical conditions
- Quasar and GRB hosting galaxies
- HII regions, post-AGB, HH, PNe
- Solar system: comets and planetary atmospheres
- Mass distribution in galaxies
- PNs in M31 and Local Group galaxies
- Stellar populations and gas kinematics
- Study of objects with high-z Ly$\alpha$ emission
- Gravitational lenses

## 2.2 SIDE as compared to other instruments

The following tables show an interesting comparison between SIDE and other existing or planned instruments in world 10 m class telescopes. Table 2 concerns the MOS mode, and Table 3 and Table 4 concern the Integral Field Spectroscopy modes (i.e. SIFU and mIFU modes).
SIDE has the additional and unique observing mode of "simultaneity", which means that both the optical and the near-IR wavelength range can be observed at the same time.

Table 2. Comparison with other instruments (MOS mode).

| Telescope | Apert. (m) | Instrument | Slit/fiber | IFU | Resolution ($\lambda/\Delta\lambda$) | Spectral Range (nm) | f.o.v (sq.deg) | $N_{obj}$ | $\eta^1$ |
|---|---|---|---|---|---|---|---|---|---|
| Keck | 10 | DEIMOS | slit | no | 1700-4800 | 410-1000 | 0.023 | 100 | 1.0 |
| VLT | 8.2 | VIMOS | slit | yes | 180-2500 | 370-1000 | 0.062 | 750 | 1.87 |
| VLT | 8.2 | FLAMES | fiber | yes | 7500-47000 | 370-1400 | 0.14 | 132 | 0.15 |
| SUBARU | 8.3 | FMOS | fiber | no | 1500-3000 | 900-1800 | 0.2 | 400 | 0.32 |
| SUBARU | 8.3 | WFMOS | fiber | no | 1000-30000 | 390-1000 | 1.77 | 3000 | 0.27 |
| GTC | 10.4 | SIDE | fiber | yes | 1500-30000 | 350-1700 | 0.09 | 1000 | 2.76 |

[1] Survey efficiency relative to DEIMOS, defined as $(Apert)^2 N_{obj} / FOV$

Table 3. Comparison with other instruments (SIFU and mIFU modes). Code: 1) Image slicers, 2) Fibers coupled to lenses, 3) Lenslet arrays. Normal text: working in the visible, **Bold text**: working in the visible and NIR.

| Telescope | Aper. (m) | Instrument | Nr. Slices, fibers or lenses | Spati. Samp. Slice/Lense/Fiber (arcsec) | FOV | $\lambda$ range ($\mu$m) | R=$\lambda/\Delta\lambda$ |
|---|---|---|---|---|---|---|---|
| **1) VLT** | **8.2** | **SINFONI/ SPIFFI(AO)** | **32** | **0.25, 0.1 or 0.025** | **8"x8"** | **1-2.5** | **1500 to 4000** |
| **1) GEMINI S** | **8.1** | **GNIRS-IFU (with AO??)** | **21 or 26** | **0.15 or 0.04** | **3.5"x4.8"** | **0.9-2.5 (5.5?)** | **1700-5900** |
| **1)GEMINI N** | **8.1** | **NIFS (AO)** | **29** | **0.1** | **3.0"x3.0"** | **0.9-2.4** | **4990-6040** |
| 2) VLT | 8.2 | VIMOS-IFU | 6400 | 0.33 or 0.67 | 54"x54" | 0.36-1.0 | 180-3100 |
| 2) VLT | 8.2 | GIRAFE/ FLAMES | 300 (15x20 or 300) | IFU mode (0.52) ARGUS 0.52 or 0.3 | 15 x 2"x3" 11.5"x7.3" | 0.37-0.9 | 11000-39000 |
| 2) GEMINI N AND S | 8.1 | GMOS | 1000 (obj) 500 (sky) | 0.2 | 5.7"x3.5" (obj) 5"x3.5" (sky) | 0.1-1.1 | 640-4400 |
| 3) SUBARU | 8.2 | Kyoto-3DII | 37x37 | 0.093 | 3.4"x3.4" | 0.4-0.9 | ~1200 |
| **3) KECK** | **10** | **OSIRIS(AO)** | **3000??** | **0.02,0.035,0.05,0.1** | **4.8"x6.4"** | **1-2.5** | **~3500** |
| **GTC** | **10.4** | **SIDE SIFU** | **2500** | **0.5** | **30"x30" 3"x3" 1.5"x1.5"** | **0.35-1.7** | **1500-30000** |

Table 4. Comparison with other instruments (SIFU and mIFU modes). Code: 1) Image slicers, 2) Fibers coupled to lenses.
Normal text: working in the visible, Bold text: working in the visible and NIR.

| Telescope | Aper. (m) | Instrument | Nr slices, fibers or lenses | Spati. Samp. slice/lense/fiber (arcsec) | Maximum FOV | λ range (μm) | R=λ/Δλ |
|---|---|---|---|---|---|---|---|
| **1) GTC** | **10.4** | **FRIDA** | **18** | **0.036 or 0.060** | **1.2"x2.2"** | **0.9-2.5** | **500, 5000, 3000** |
| 1) VLT | 8.2 | KMOS | 24 IFUs 14 slices/IFU | 0.2 | <u>24 x 2.8"x2.8"</u> | 1-2.45 | 3400-3800 |
| 1) VLT | 8.2 | MUSE (AO) | 24 IFUs 12 Slices/IFU? | 0.025 or 0.2 | 1'x1' | 0.46-0.93 | 2000-4000 |
| 1) Various | ~8 | SWIFT | 44 | 0.1 or 0.15 | 6.6"x13.6" | I/Z 0.65-1.0 | 3500-8500 |
| **1) JWST** | **6.5** | **NIRSpec** | **30** | **0.1** | **3"x3"** | **0.6-5** | **100, 1000, 2700** |
| **1) JWST** | **6.5** | **MIRI** | **4 IFUs? ~10 Slices/IFU** | **0.27 to 0.76 Depending on channel** | **3"x3.9" to 6.7"x7.7" depnding on channel** | **5-27** | **2400-3700** |
| 2) HET | 9.2 | VIRUS | 132 IFUs 247/IFU | 1 | <u>145x2.8"x2.8"</u> | 0.34-0.57 | 5000 opt to NIR |
| 2) LBT | 2x8.4 | LUCIFER (2 inst.) | 682 | 0.025 or 0.25 | ??????? | NIR | ? |
| GTC | 10.4 | SIDE | 2500 | 0.5 | 30"x30" 3"x3" 1.5"x1.5" | 0.35-1.7 | 1500-30000 |

## 3. INSTRUMENT MAIN SYSTEMS

### 3.1 Instrument layout

SIDE is planned to be formed by a set of 11 spectrographs, of which 10 are dedicated to low resolution and one to high resolution. All these spectrographs are fed in parallel by the fibers coming from the MOS units, the SIFU and the MiniIFUs. Two focal stations are planned to be used: Nasmyth (Nas) and Folded Cassegrain (F-Cass). The MOS units are configured at Nas by a Fiber Positioner, while the SIFU and MiniIFUs stay at F-Cass, the MiniIFUs being positioned by a second (and smaller) fiber positioner. All the 11 spectrographs should be contained in a dedicated room with controlled temperature, which could be placed below the Nas platform, in order to minimize the fiber cable length. Both focal stations should be provided of a field corrector optical system.

### 3.2 The field correctors

The field corrector optics is a large assembly formed by two lenses of about one meter in diameter in the case of the Nasmyth focal station, and about 50 cm across in the case of the F-Cass. A possibility has to be explored to avoid a field corrector at Nasmyth, in case the fiber positioner can cope with the original focal plane of the telescope.

### 3.3 The fiber positioners

The fiber positioner for the MOS units (Nas focal station) follows the concept developed by the Berkeley Laboratory (LBNL). The LBNL design divides the focal plane into ~1000 hex cells. Each fiber is individually-actuated within one hex cell. The fiber motion extends slightly beyond its hex cell, such that there is no spot on the focal plane that cannot be reached by at least one fiber. The downside of the LBNL design is that the fibers cannot be densely packed in any one part of the focal plane. For example, one could not pack all of the fibers into 1/4 of the focal plane if there were a compact cluster of targets, or even follow the large-scale structure of clustering for galaxies. This limitation can be mitigated by a clever set of offset pointings, made possible by the fast reconfiguration time (discussed later). The upside of the LBNL design is that the reconfiguration time is extremely fast. The entire focal plane can be reconfigured in under 20 sec. The local precision need be only 1 part in 1000 for each actuator to achieve 30 micron precision, which is possible running open loop. There is also an inherent robustness to this design: any mechanical failures would compromise only 1 MOS science unit.

The focal plane is defined by a 2-inch thick aluminum block machined to the exact shape of the focal surface. The active focal plane is 99.35 cm to cover the 20' field, but the block would extend beyond that by several cm for mounting purposes. 1003 circular holes are drilled into this block with a center-to-center separation of 2.92 cm. Each of the 1003 fiber actuators is plugged into one of these holes from the backside. Each fiber actuator is mechanically independent of every other. The fiber is moved in a 2-dimensional plane, since the focal surface is approximated as flat within the footprint of one fiber. There are two MicroMo 8mm motors that control the two degrees of freedom. There is a geared, linear rack on the top of the actuator that moves the fiber radially from the center. The entire top assembly (in blue in Fig. 2) is moved rotationally by a second MicroMo motor, which points downward to a small gear on the bottom of the assembly. That gear drives against a large gear that is fixed to the actuator outer casing (in green). The fiber reach is 1.686 cm from its center position,

The fiber (in red) is mounted in the end of the rack in a bushing. A serious consideration for any fiber-fed spectrograph is that the light is only telecentric at the center of the field, and will in general deviate by several degrees at the edge of the field. The LBNL design does allow for the light to approach the focal plane at an angle, with the reducing lens mounted on each fiber compensating. Since each fiber is local to its position in the focal plane, the correct lens can be manufactured for each fiber. The fiber + lens assembly does need to be prevented from rotating with the azimuthal gear. This is achieved by holding the fiber in a flexible hyperdermic tubing that is fixed on the backside of the actuator. The fiber + lens then rotates within the bushing in the rack on the top end, and maintains its position rotationally relative to the sky.

The actuators are run in open loop with a positional accuracy of ±30 μm (corresponding to ±0.04 arcsec on the focal plane). An initial position reference is supplied by an index switch of two gold-plated pins in a crossbar configuration. One pin (actually a wire) has flexibility so that the switches can slightly over-travel and not change the zero-point by over-stressing the crossbars. The rotational movements are supported by ball bearings ranging in size from 3 mm to 28 mm diameter. An electronics assembly sits on the back side of each actuator. See Fig. 2 for the details.

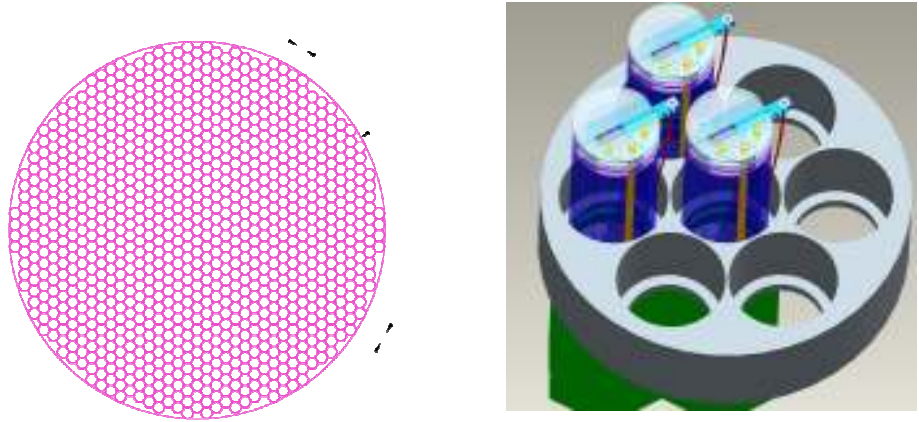

Fig. 2. (Left) Focal plane divided into 1003 hex cells, with a center-to-center spacing of 2.92 cm. An LBNL fiber actuator with one fiber is mounted in each cell. The fibers can reach beyond the hex cell, such that every position on the focal plane is accessible by at least one fiber. (Right) Top-view of several LBNL fiber actuators. The fiber itself is shown in red. Radial motion of the fiber is controlled by the rack-and-pinion.

The fiber positioner for the MiniIFUs (F-Cass) follows the design developed by S. Eikenberry at the University of Florida. The primary structural component of the mechanism is a top plate, to which all probe arm and motor components are mounted. The plate has a central cutout allowing light in the converging beam from GTC to fill the unvignetted region at the focal plane. We assume that the "home position" of the optical fibers is a ring just outside the focal plane circle. Each mIFU fiber bundle is permanently attached to the end of its own positioning arm and the 27 positioner arms each patrol a "slice of pie" region just above ("upstream" of) this focal plane. In operation, each arm tip will move to the desired location for the mIFU target.

Each arm will be positioned via two stepper-motor-driven actuators. The dedicated arms for each mIFU create the possibility of efficient "on the fly" repositioning of SIDE fiber setups during the night's observations.

### 3.4 The Dual VIS-NIR spectrograph

SIDE will use two kinds of spectrograph: a low resolution VIS-NIR device (hereafter called "Dual") and a high resolution VIS device. The Dual device will actually be a set of ~10 identical units working in parallel. Each spectrograph unit will have a mirror collimator, a dichroic with a 950 nm cutoff and two 6-lens cameras for the VIS and NIR respectively, following the same baseline as the Sloan Digital Sky Survey spectrograph. The initial optical requirements for each of the Dual spectrograph are to cover the spectral range from 400 nm to 1700 nm with 4 exposures at R=4000 for the VIS arm and 3 exposures for the NIR arm (from 950 nm to 1700 nm) and almost the entire spectral range from 525 nm to 1700 nm in one single exposure at lower resolution R=1500. The spectral range from 1400 to 1700 nm still needs to be assessed as some cooling of the whole spectrograph would be necessary, and this would imply a large cost increase. Fig. 3 shows the basic optical layout of this device, here with the vrism (VPH plus prism) setup for the R = 1500 resolution. Fig. 4 shows a view of the 3-D setup.

All the spectrographs would be fed by a double pseudoslit in order to optimize the number of objects collected by each. Simulations showed that the offset slit and the increased occultation due to the double pseudolist has negligible effect on the image quality and efficiency of the spectrograph.

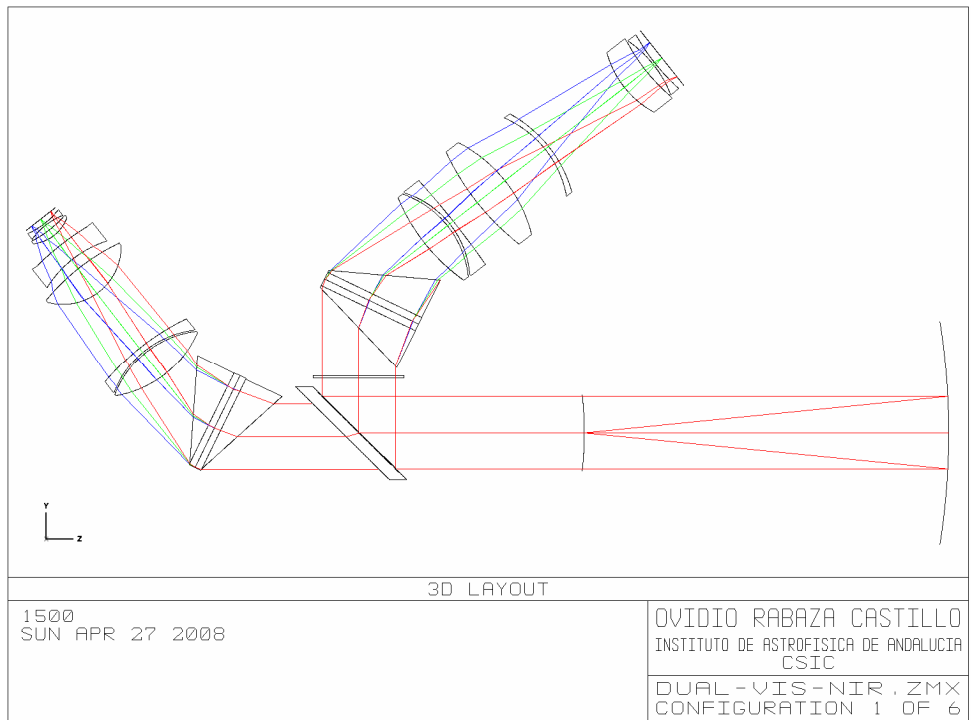

Fig. 3. Optical layout of the low resolution VIS-NIR spectrograph for resolution R=1500.

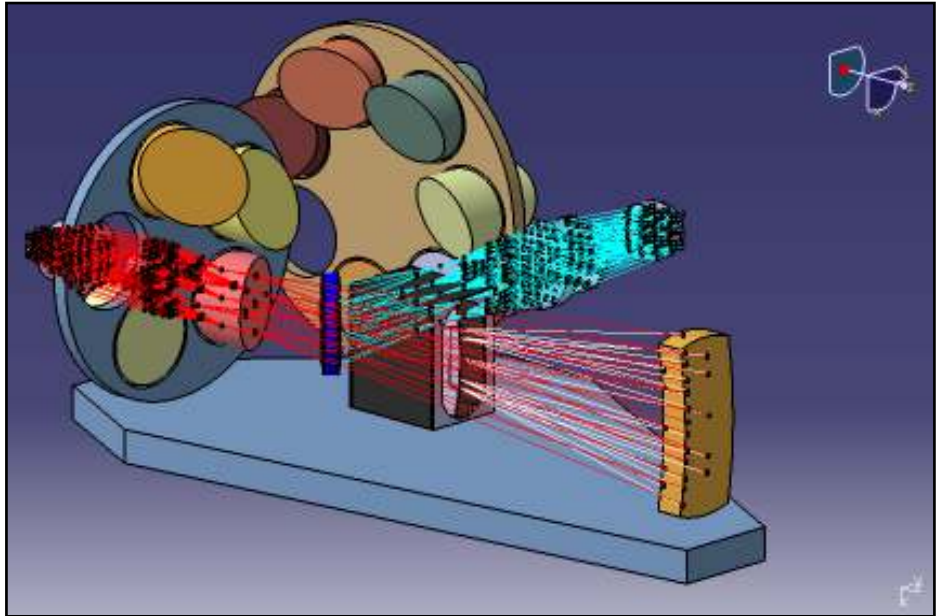

Fig. 4. Optomechanical view of the low resolution spectrograph.

### 3.5 The Hi-Res VIS spectrograph

The high resolution spectrograph is a much larger device and should be one single piece of equipment. It should be fed by some of the MOS units (about 90 units) and the MiniIFUs. It is planned to provide resolutions from R=7900 to R=29000. It will also have a double pseudoslit in order to accommodate the large number of fibers. The required resolution could be attained with a collimated beam diameter of about 300 mm. We adopted that value and assumed that the dispersion would be provided by a large reflection grating, working in 1st interior order. The collimator is quite simple and effective. The only optical element needed is a spherical mirror whose radius is twice the focal length. The pseudo-slit must be made in the form of a thin "fan," whose individual fibers point toward the mirror with their individual axes all pointing back such that they intersect an optical axis at the mirror's center of curvature. The collimator f-ratio that best accomplished the spot diameter required was f/7, with a focal length of 2100.0 mm and a spherical mirror radius of 4200.0 mm.

In order to cover all the spectral range, the grating tilt will be tunable and will span 23.3º, from 36º to 59.3º of angle from the collimator. The advantage of such setup is the use of one single grating without the need of refocusing. On the other hand, the grating is a difficult and expensive optical part. The grating will surely have to be made as a mosaic.

These parameters, led to a required camera focal length of about 880.0 mm and a camera field radius of about 2.85 degrees, which covers the corners of the array. The dispersion values required to accommodate central wavelengths in the (0.37 to 0.85)-micron range, with resolution in the R = (7,900 to 29,000) range, appear to be attainable with a single 1st-order reflection grating as described above but it will require a (310 x 600)-mm grating so as to accommodate the 0.85-micron central wavelength set up without vignetting at the ends of the grating. Refracting cameras of such dimensions and spectral characteristics are not feasible so a catadioptric camera will be required.

Fig. 5 shows a layout of the preliminary Hi-Res VIS spectrograph design. We conclude that image slicing, by use of fiber coupling with units containing 7 fibers, makes it possible to achieve spectral resolution and wavelength coverage mandated by the science goals for SIDE, with collimated beam sizes that are buildable and affordable.

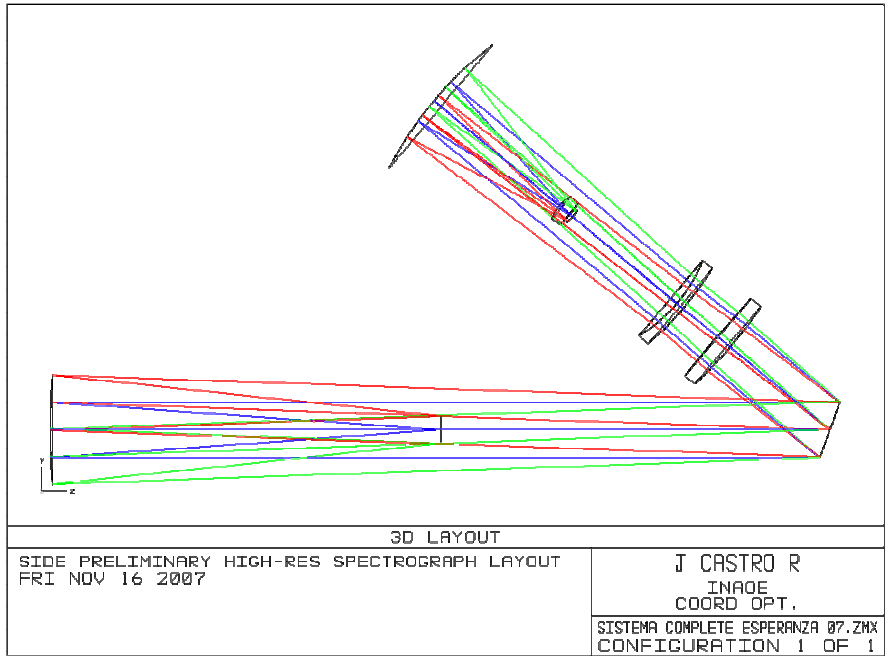
Fig. 5. Optical layout of the High Resolution spectrograph.

## 4. CONCLUSIONS

If finally approved and built, SIDE will be an outstanding instrument, which will remain competitive for many years and will provide first class data, especially valuable for large survey programs. It will provide the GTC community with new and unique observing capabilities, and particularly will lead the survey spectroscopy for more than a decade.